\documentclass[twocolumn,showpacs,amsmath,prl]{revtex4-1}

\usepackage{color}
\usepackage{graphicx}
\usepackage{epsfig,psfrag}
\usepackage{amsmath}
\usepackage{amssymb}
\usepackage{amsbsy}
\usepackage{amsthm}
\usepackage{amsfonts}
\usepackage{wasysym}
\usepackage{bbm}
\usepackage{tabularx}
\usepackage{euscript}
\usepackage{color}
\usepackage{enumerate}
\usepackage{amsfonts}
\usepackage{exscale}
\usepackage{bbold}
\usepackage{float}

\usepackage[colorlinks,citecolor=blue]{hyperref}

\renewcommand{\v}[1]{{\boldsymbol{#1}}}

\newcommand{\be}{\begin{eqnarray}}
\newcommand{\ee}{\end{eqnarray}}

\newcommand{\Eq}[1]{equation~(\ref{#1})}
\newcommand{\Fig}[1]{Fig.~\ref{#1}}

\newcommand{\<}{\langle}
\renewcommand{\>}{\rangle}

\newcommand{\Tr}{{\rm Tr}}

\newcommand{\ra}{\rightarrow}

\def\bea{\begin{eqnarray}}
\def\eea{\end{eqnarray}}
\def\bra#1{\left\langle#1\right|}
\def\ket#1{\left|#1\right\rangle}
\def\avg#1{\left\langle#1\right\rangle}

\def\Tr{\mathrm{Tr}}

\def\Eq#1{Eq.~(\ref{#1})}
\def\Fig#1{Fig.~\ref{#1}}

\begin{document}

\title{Superconductivity in the presence of strong electron-phonon interactions and frustrated charge order}
\author{Zi-Xiang Li$^{1,2}$, Marvin L. Cohen$^{1,2}$,}
\author{Dung-Hai Lee$^{1,2}$}\email{Corresponding author: dunghai@berkeley.edu}
\affiliation{
$^1$ Department of Physics, University of California, Berkeley, CA 94720, USA.\\
$^2$ Materials Sciences Division, Lawrence Berkeley National Laboratory, Berkeley, CA 94720, USA.
}

\begin{abstract}
We study the superconductivity of strongly coupled electron-phonon systems where the geometry of the lattice frustrates the charge order by the sign-problem-free Quantum Monte Carlo(QMC) method. The results suggest that with charge order frustrated, the superconductivity can benefit from strong electron-phonon interaction in a wide range of coupling strengths.
\end{abstract}

\maketitle

\emph{Introduction}.-Strong electron-phonon (e-ph) interactions are often regarded as being beneficial for strong Cooper pairing and high-temperature superconductivity (SC). However, as the e-ph interaction gets strong other electronic or lattice instabilities often set in to preempt the high-temperature superconductivity. This problem has been appreciated in the literature for a long time \cite{CohenAnderson, Varma,Rice,Scalapino,White,Fehske,Meyer,Alexandrov,Bauer,Kivelsonbook,Fehskebook}. It has received renewed interest recently, and sign-problem-free quantum Monte Carlo (QMC) simulation has been used to study it\cite{Kivelson}.\\

For the Holstein model on a square lattice, Ref.\cite{Kivelson} shows  that as the e-ph interaction gets strong, the $\v Q=(\pi,\pi)$ charge density wave (CDW) susceptibility surpasses that of SC. Moreover, when this happens, the Migdal-Eliashberg (ME) theory fails. Interestingly, this failure occurs when the control parameter of the ME theory, namely, the ratio of the phonon energy to the electron Fermi energy,  is small. In addition, it is pointed out that the CDW wavevector is unrelated to Fermi surface nesting. (The electron hopping includes next neighbors, hence the Fermi surface is not nested, even at half-filling.) Physically when the e-ph interaction becomes sufficiently strong, bipolarons form.
Under such conditions the ground state consists of bipolarons occupying one of the sublattices so that each doubly occupied site is surrounded by empty ones. In such an arrangement electrons on the doubly occupied sites can virtually hop to the neighboring sites to gain the kinetic energy. This is similar to the super exchange mechanism of repulsive systems. \\

 Here we ask the question ``what if the CDW is frustrated by the geometry of the lattice, will this frustration allow SC to benefit from stronger e-ph interactions.'' To answer this question we study the Holstein model on both square (un-frustrated) and triangular (frustrated) lattices.  Following Ref.\cite{Kivelson} we quantify the strength of the e-ph interaction by the dimensionless parameter  $\lambda =\alpha^2\rho(E_F)/K$, where $\rho(E_F)$ is the density of states at the Fermi energy and $\alpha$ is the e-ph interaction parameter (see. \Eq{model}).
The main results summarized in the following are obtained for $ 0.2\le\lambda\le 0.8$ and $\hbar\omega/E_F=0.1$ and $0.3$ at different electron densities and temperatures. (1) For the half-filled square lattice at $T=0$ we find no evidence for SC order for the lattice sizes we have studied. On the other hand, we do find $\v Q=(\pi,\pi)$ CDW order for the entire range of $\lambda$ we studied. (2) For a half-filled triangular lattice at $T=0$, on the other hand, we  find SC order in the entire range of $\lambda$. In contrast, the CDW order (with $\v Q=(\pm 4\pi/3,0)$) only exists for $ 0.4 \lesssim\lambda\le 0.8 $ when $\hbar\omega/E_F=0.1$, and for $ 0.6\lesssim\lambda\le 0.8$ when $\hbar\omega/E_F=0.3$. In these more restricted ranges of $\lambda$, SC and CDW coexist. (3) For the half-filled triangular lattice with $\lambda=0.4$ (where SC is strongest) we determine the Kosterlitz phase transition temperature to be $T_c\approx t/10$. (4) In the SC-CDW coexistence phase at, e.g., $\hbar\omega/E_F=0.1$,
charge fluctuations are significantly stronger on one  sublattice of the tri-partite triangular lattice (see discussion below). Moreover, the SC order parameter is the strongest on this sublattice. Through these sites SC can survive at strong e-ph coupling even after the CDW order has set in. (5) At half filling  a single-particle gap exists in the non-SC phase for both square and triangular lattices. (6) For triangular and square lattices doped away from half filling we find the low temperature CDW/SC susceptibilities are suppressed/enhanced relative to  those at half filling.\\

The results summarized above  make the case that the frustration of CDW  allows SC to benefit from stronger e-ph coupling without being preempted by the charge order. \\

\begin{figure}[t]
\includegraphics[height=1.45in]{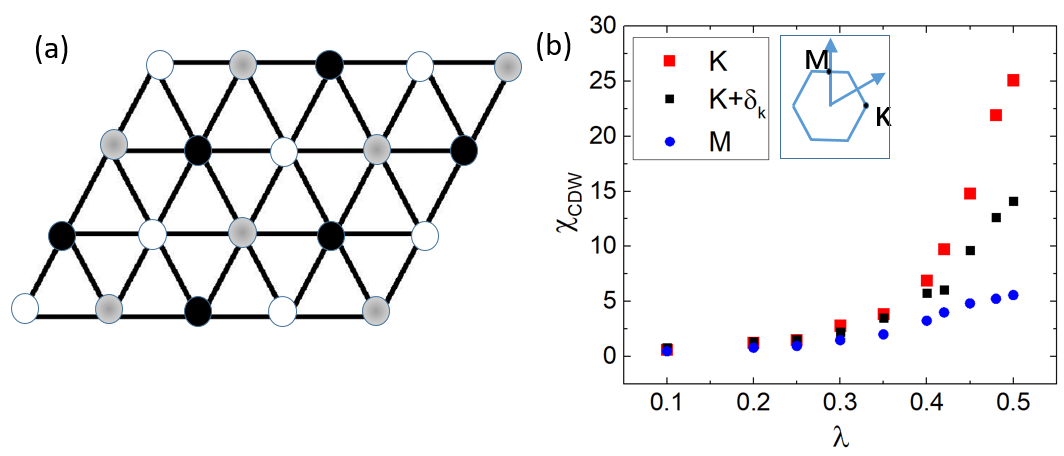}
\caption{ (a) A schematic figure of the  $\sqrt{3}\times\sqrt{3}$ CDW ordered state in the strong e-ph coupling limit on triangular lattice. The ordering wavevectors are  $(\pm 4\pi/3,0)$. The black circles represent the sites with $\<n_i\> >1$ and the white circles represent sites with $\<n_i\><1$.  The gray circles stand for sites where large charge fluctuations and $\<n_i\>\approx1$.   (b) The CDW susceptibility at different momenta for the triangular lattice, where $\hbar\omega/E_F$ is set to $0.3$ and the temperature is set to $t/16$. Here $\v K=(4\pi/3,0)$, and $\v M=(0,2\pi/\sqrt{3})$ and  $
\v K+\delta_\v k$ is a momentum closest to $\v K$ on a lattice with linear dimension $L=12$. }
\label{fig1}
\end{figure}

Before discussing the details we present a physical picture which enables one to understand the above results.
When $\lambda$ becomes sufficiently strong bipolarons form. In the charge ordered phase the bipolarons are localized.
For the square lattice, which is bipartite, the bipolarons localize on one of the sublattices so that virtual hopping can lower the kinetic energy. For the triangular lattice, however, such an arrangement is impossible. This is the same as the frustration encountered in the
antiferromagnetic Ising model on a triangular lattice. This obstruction toward charge order benefits SC. To understand the coexistence phase we note that the ground state of the AF Ising model is macroscopically degnerate ($exp(cN)$ ($(c\sim O(1))$) spin patterns have the same energy)\cite{Stephenson}. Moreover, it has been shown that out of these degenerate spin patterns, a class of $\sqrt{3}\times\sqrt{3}$  spin configurations (characterized by $\v Q=(\pm \frac{4\pi}{3},0)$ wavevectors)  are stabilized at low temperatures due to an ``order by disorder'' mechanism \cite{Sondhi-2000,Sondhi-2001,Balents-2005,Damle-2005,Troyer-2005,Prokofev-2005,Hassan-2007}. In the present problem we expect the analogous CDW pattern, with $S^z=+1\ra$ double occupancy and $S^z=-1\ra $ single occupancy (see \Fig{fig1}(a)), to be stabilized by either thermal or the quantum fluctuations (introduced by the hopping of electrons). This expectation is supported  by the simulation result -- the strongest CDW susceptibility is associated with wavevector $(\pm \frac{4\pi}{3},0)$  as shown in \Fig{fig1}(b). The same order by disorder mechanism predicts charge fluctuation to be significantly stronger on one of the three $\sqrt{3}\times \sqrt{3}$ sublattices. Each site in this sublattice is surrounded by a hexagon of sites where the charge density alternates between $\<n_i\>>1$ and $\<n_i\><1$. These sites are analogous to the ``flippable sites'' in the entropy-stabilized $\sqrt{3}\times \sqrt{3}$ pattern  of the antiferromagnetic Ising model (spins on the flippable sites are surrounded by alternating spin ups and spin downs, hence they feel no internal field). Through these large charge fluctuation sites, SC can survive even after the CDW order has set in.
\\

\noindent{{\bf The model}}\\
In the following discussions we consider the Holstein model defined on both square and triangular lattices $H = H_e + H_p + H_{ep}$ where
\be
H_e &=& - \sum_{\avg{ij},\sigma} t_{ij} (\psi^\dagger_{i,\sigma}\psi_{i,\sigma} + h.c) - \mu \sum_i \hat{n}_{i,\sigma} \nonumber\\
H_p &=& \sum_{i} (\frac{\hat{P^2_i}}{2M} + {K\over 2}  \hat{X}_i^2); ~~H_{ep} = \alpha \sum_{i} \hat{n}_i \hat{X}_i
\label{model}
\ee
Here, $\psi_{i,\sigma}$ annihilates an electron with spin polarization $\sigma$ on lattice site $i$, $\mu$ is the chemical potential, and $\hat{n}_i$ is the electron number operator associated with site $i$. For the triangular lattice we set the hopping integrals $t_{ij}$ to $1$ between nearest neighbors. For the square lattice we set the nearest-neightbor hopping integral to $1$ and second neighbor hopping integral to $-0.2$ to avoid a nested Fermi surface at half-filling. In the rest of the paper we use $t$ to denote the nearest neighbor hopping matrix element for both triangular and square lattices. $H_p$ describes a dispersionless Einstein phonon with frequency $\omega=\sqrt{K/M}$, where $\hat{X}_i$ is the phonon displacement operator and $\hat{P}_i$ is its conjugate momentum. $H_{ep}$ describes the e-ph coupling with $\alpha$ being the coupling constant. As mentioned earlier, the e-ph coupling strength is characterized using the dimensionless parameter  $\lambda =\alpha^2\rho(E_F)/K$. Moreover we set $\hbar\omega/E_F$ to be either 0.1 or 0.3 in the QMC studies.

Due to  the presence of time reversal symmetry and particle number conservation in the electronic part of the action for arbitrary phonon configurations the partition function of \Eq{model} is free of the fermion minus sign problem where it is subjected to determinant QMC  simulation\cite{Assaad-2005,Congjun,Yao1,Yao2,Xiang,Wang,Yao3}. In the literature many QMC simulations have been applied to the Holstein model on non-frustrated lattices\cite{Scalapino2,Assaad-2007,Hohenadler,Devereaux,Johnston,Kivelson,Weber,Scalettar1,Ziyang1,Scalettar2,Ziyang2,Kivelson2}.  Here the introduction of lattice frustration is a new aspect. We preform zero and non-zero temperature  QMC simulations. Due to the absence of the sign problem our simulation can be carried out at any temperature for large system sizes.  The  details of the QMC simulations can be found in the Supplementary Materials.  \\

\begin{figure}[t]
\includegraphics[height=2.65in]{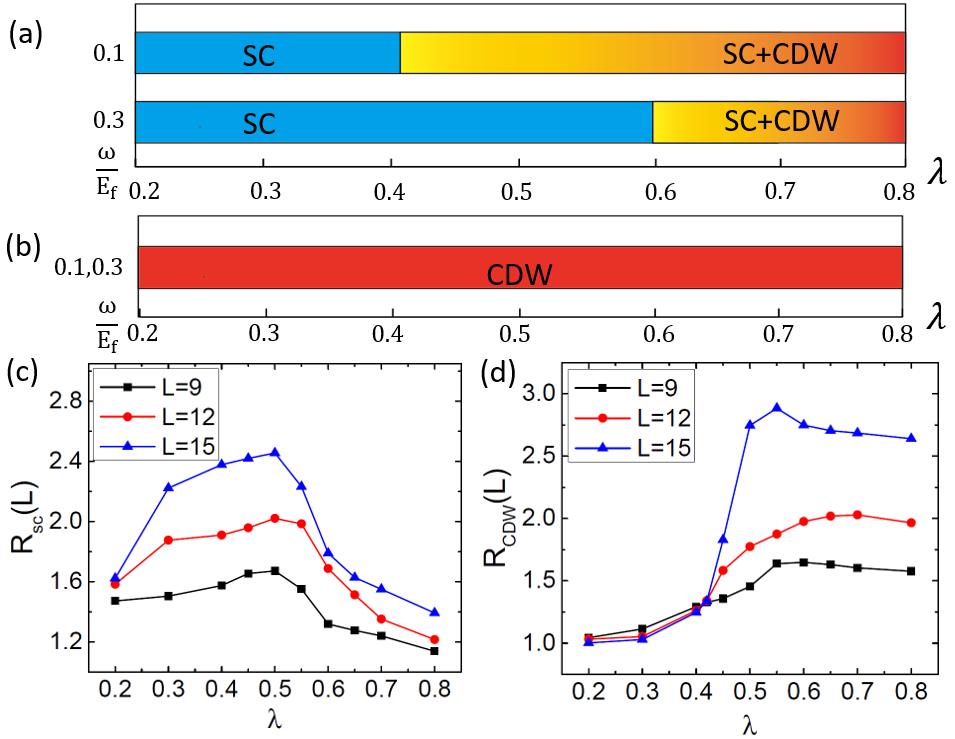}
\caption{ (a) The zero-temperature phase diagram of the Holstein model in the e-ph coupling range $0.2\le\lambda\le 0.8$ and $\hbar\omega/E_F=0.1,0.3$ for the triangular lattice. For both values of $\hbar\omega/E_F$ the ground state possesses SC order in the entire range of $\lambda$. However, for $\hbar\omega/E_F=0.1$ the  CDW order sets in to coexist with SC for $0.4\lesssim \lambda\le 0.8$. For $\hbar\omega/E_F=0.3$ the CDW sets in for $0.6\lesssim\lambda\le 0.8$. (b) The zero temperature phase diagram for the square lattice with $\hbar\omega/E_F= 0.1,0.3$ and $0.2\le\lambda\le 0.8$. Here we only find the $\v Q=(\pi,\pi)$ CDW order. (c) The RG invariant ratio, $R_{SC}(L)$, for the SC order at $\hbar\omega/E_F=0.1$ on half-filled triangular lattice. The result indicates SC long-range order. (d) The RG invariant ratio, $R_{CDW}(L)$, for the CDW order at $\hbar\omega/E_F=0.1$ on half-filled triangular lattice. The result is indicative of CDW disorder-order phase transition at $\lambda \approx 0.42$.  }
\label{fig2}
\end{figure}

\noindent{{\bf Zero-temperature and half-filling}}\\
{\bf I. The phase diagram} \\
We employ projector QMC to study the ground state of \Eq{model} for $\hbar\omega/E_F=0.1,0.3$ and $0.2\le\lambda\le 0.8$ at zero temperature. Through a finite-size scaling analysis for systems with linear dimension $L=6,9,12,15$ (the details can be found in the supplementary materials) we obtain the zero-temperature phase diagrams in the specified range of $\lambda$ as shown in \Fig{fig2}(a),(b). For the triangular lattice (\Fig{fig2}(a)) SC long-range order exists in the entire range of $\lambda$ we studied. Moreover, this is true for both $\hbar\omega/E_F=0.1$ and $0.3$. However, for $\hbar\omega/E_F=0.1$ CDW sets in to coexist with SC for $0.4\lesssim\lambda\le 0.8$. For $\hbar\omega/E_F=0.3$, CDW order is weakened and but it still sets in to coexist with SC for $0.6\lesssim\lambda\le 0.8$. The CDW ordering wavevectors are $\v Q=(\pm 4\pi/3,0)$ and a schematic figure of it is given in \Fig{fig1}(a). In contrast, for the square lattice there is no SC order (at least within the lattice sizes we studied.) Instead we find CDW order in the entire $\lambda$ range we studied (see \Fig{fig2}(b)). The ordering wavevector is  $\v Q=(\pi,\pi)$.
Note that our phase diagram excludes $\lambda<0.2$. This is because for small $\lambda$, weak SC or CDW orders can be suppressed by the non-zero energy gap caused by the finite lattice size, and hence prevent us from drawing conclusions in  the thermodynamic limit.  However, we do expect the presence of SC order in the thermodynamic limit due to the standard argument that SC is the generic instability for Fermi surface possessing time reversal symmetry. Because the bandstructure does not possess  Fermi surface nesting there is no CDW instability. In \Fig{fig2}(c) and (d) we present the ``RG-invariant ratio'' $R=S(\v Q)/S(\v Q+\delta \v q)$ as a function of system size $L$ for the SC and CDW orders. Here $S(\v Q)$ is the Fourier transform of the SC/CDW correlation functions, and $\v Q=(0,0)$ for SC, and $\v Q=(\pi,\pi)$ or $(\pm 3\pi/4,0)$ for the CDW on the square and triangular lattices.  $\delta\v q$ is a small wavevector introduced to enable a comparison between the correlation function at the expected ordering wavevector and a wavevector nearby. Long-range order implies the divergence of $R$ as $L\rightarrow \infty$, while short range order means $R\rightarrow 1$.  The results clearly support the phase diagram presented in \Fig{fig2}(a). Compare the results for the square and triangular lattice we conclude that frustration of the charge order enables the SC to prevail for a much wider range of strong e-ph interaction.\\

\begin{figure}[t]
\includegraphics[scale=0.43]{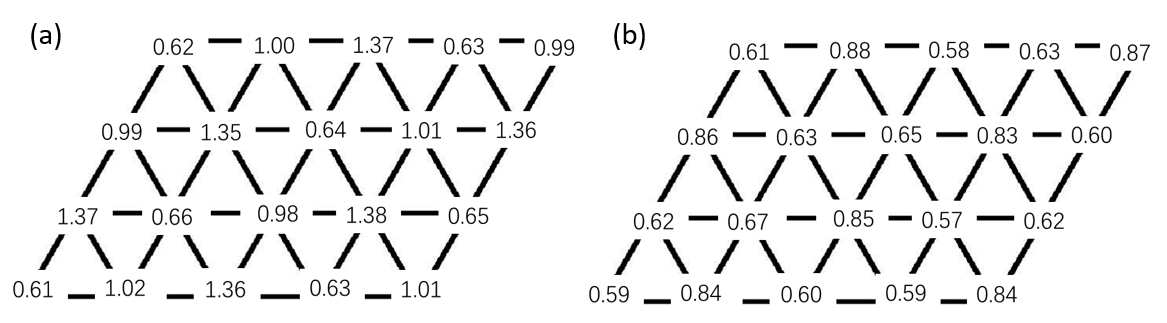}
\caption{The local density (a) and its mean square fluctuations (b) in the CDW-SC coexistent phase on the triangular lattice at $\hbar\omega/E_F = 0.1$. The calculation is carried out at zero temperature for $L=12$ and $\lambda=0.5$.  }
\label{fig3}
\end{figure}

\noindent{{\bf II. The coexistence phase}}\\
In order to gain more insight into the SC/CDW coexistence phase, we turn on a tiny pinning potential consistent with the periodicity of the CDW.
We then compute the expectation value of local electron density  $n_i = \avg{c^\dagger_i c_i}$ and its mean square fluctuation  $\Delta n_i^2 = \avg{\hat{n}_i^2}-\avg{\hat{n}_i}^2$.
The details of the calculation can be found in the supplementary materials. The result for the electron density distribution is shown in \Fig{fig3}(a), which  clearly reveals the $\sqrt{3}\times\sqrt{3}$ periodicity. The results for $\Delta n^2$ is presented in \Fig{fig3}(b). It shows a significantly  stronger charge fluctuation on the lattice sites with $\avg{n_i}\approx 1$. 
In addition, we have also computed the SC correlation function in the coexistence phase. Remarkably, the correlation is significantly stronger among the sites with larger charge fluctuation. These results suggest the SC coherence within the CDW is enabled by the  ``flippable'' sites,  which in turn is caused by the geometric frustration.\\

\begin{figure}[t]
\includegraphics[scale=0.52]{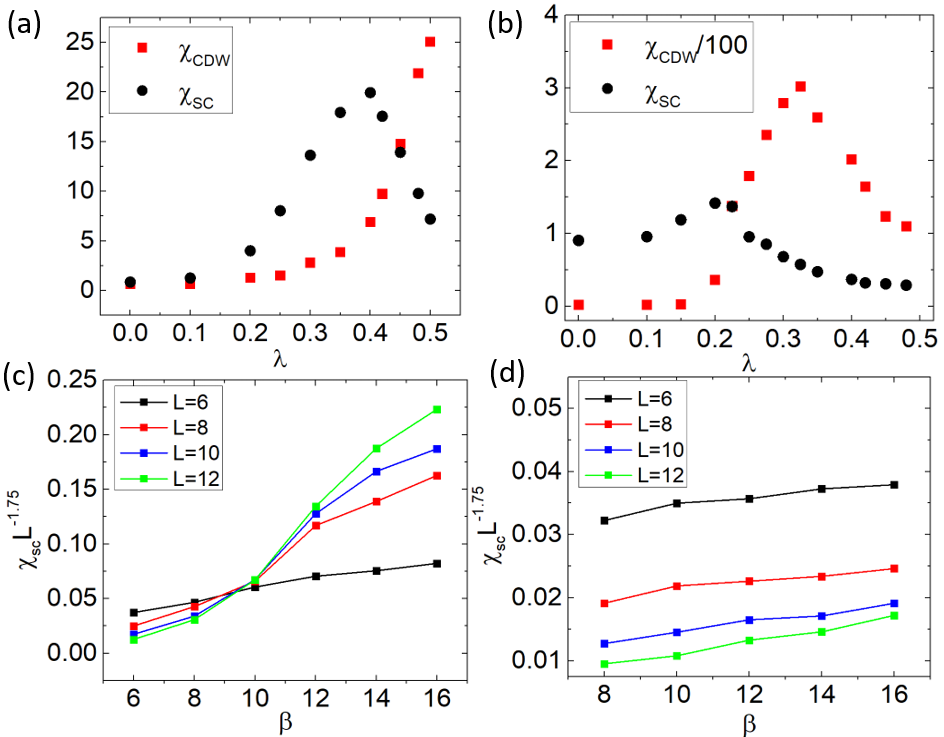}
\caption{ (a) The SC and CDW susceptibilities on the triangular lattice for $\hbar\omega/E_F=0.3$. The wavevectors of the SC and CDW are $(0,0)$ and $(4\pi/3,0)$, respectively. (b) The SC and CDW susceptibilities for $\hbar\omega/E_F=0.3$ on a square lattice. The wavevectors of the SC/CDW are $(0,0)$ and $(\pi,\pi)$, respectively. The results in (a) and (b) are obtained at   $T = t/16$ and $L=12$. (c) The scaled SC susceptibility, $\chi_{SC}L^{-2-\eta_c}$ where $\eta_c=0.25$, for the triangular lattice. The e-ph coupling is set to $\lambda=0.4$. Here $\beta$ is inverse temperature in unit of $1/t$. The crossing point indicates the Kosterlitz-Thouless transition temperature: $T_c \approx t/10$. (d) The same plot for the square lattice, here $\lambda$ is set to 0.2. The absence of the crossing and the fact that $\chi_{SC}L^{-2-\eta_c}$ decreases with increasing $L$ implies that, if it exists, the SC transition temperature is below the lowest temperature we calculated. The results in (c) and (d) are obtained for $L=12$ and $\hbar\omega/E_F = 0.3$.
}
\label{fig4}
\end{figure}

\noindent{{\bf Half-filling and non-zero temperatures}}\\
\noindent{{\bf I. The SC and CDW susceptibilities}}\\
Next we fix the temperature and linear lattice size to $T=t/16$ and $L=12$ and compute the SC and CDW susceptibilities as a function of $\lambda\in [0.0,0.5]$. For the triangular lattice, the CDW susceptibility peaks at wavevector $\v Q=(\pm 4\pi/3,0)$ as shown in \Fig{fig1}(b). Moreover as shown in \Fig{fig4}(a) the SC susceptibility is enhanced with increasing $\lambda$ till $\lambda\approx 0.4$. For larger $\lambda$ the CDW susceptibility rises which suppresses the SC susceptibility. For comparison, we also plot the CDW and SC susceptibility for the square lattice in \Fig{fig4}(b). Similar to the triangular lattice result, when CDW ordering tendency gets stronger  SC is suppressed.  Moreover, upon taking the absolute scale of the susceptibility into account, it is seen that the CDW/SC susceptibility is strongly suppressed/enhanced on the triangular lattice. \\

\noindent{{\bf II. The Kosterlitz-Thouless transition} }\\
We estimate the SC Kosterlitz-Thouless transition temperature $T_c$ through the well-known scaling behavior of the SC susceptibility ($\chi_{SC}$) at the KT transition: $\chi_{SC} \sim L^{2-\eta}$, where  $\eta = 0.25$. Upon fixing $\lambda = 0.4$ for the triangular lattice and $\lambda =0.2$ for the square lattice (these are the $\lambda$ values at which the SC susceptibility is the strongest at $T=t/16$) we plot $L^{-2+\eta}\chi_{C}$ as a function of temperature in \Fig{fig4}. The crossing of the curves for different $L$ marks the phase transition.  The result suggests that the transition temperature for  triangular lattice is $T_c\approx t/10$ (\Fig{fig4}(a)). In contrast, for the square lattice no crossing is observed for  $T\ge t/16$ (\Fig{fig4}(b)). When combined with the zero temperature result this suggests the absence of SC. This comparison again provides the evidence for frustration enhanced SC on triangular lattice.\\

\begin{figure}[t]
\includegraphics[height=1.42in]{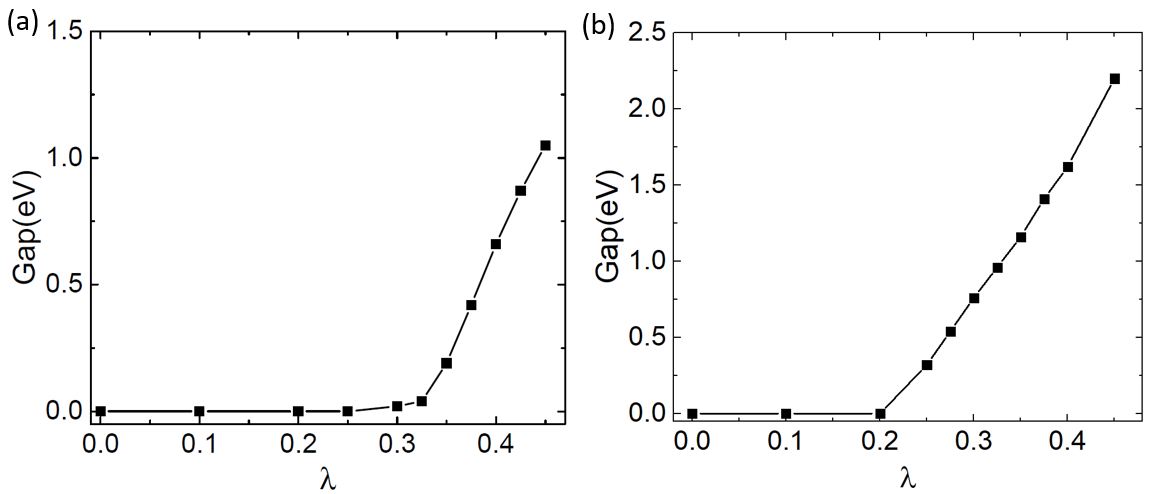}
\caption{(a) The evolution of single-particle gap, extracted from the spectral function $A(\v k_F,\omega)$, at the Fermi momentum as a function of $\lambda$ on a triangular lattice. (b) The same plot for the square lattice. The results are obtained for $\hbar\omega/E_F = 0.3$, $L=12$ and $T=t/8$. This temperature is above the highest SC transition temperature for both lattices. }
\label{fig5}
\end{figure}

\noindent{{\bf III. The pseudogap} }\\
Stimulated by the phenomenology of the cuprates, single-particle gaps above the SC transition is of considerable interests.
In the Holstein model, we expect a single-particle gap to accompany the bipolaron formation. Moreover, because the hopping of bipolarons is suppressed at large e-ph coupling, which results in a small SC phase stiffness,  we expect a pseudogap can persist above the SC transition temperature $T_c$.

We have computed the single-particle gap for both triangular and square lattices at half-filling. It is deduced from the imaginary-time single-particle Green's function  through a stochastic analytical continuation method\cite{Sandvik}.
The electron spectral  function $A(\v k_F,w)$ at different temperatures are included in the supplementary materials. As shown in \Fig{fig5}(a) and (b) the pseudogap at $T=t/8$ (which is above the highest $T_c$ for both systems) undergoes a sharp upturn around the $\lambda$ value where the CDW susceptibility rises (which signifies the bipolaron formation). In \Fig{figSM5}(d) we study the pseudogap onset temperature for the hal-filled triangular lattice. The result $T^*\approx 0.45 t$ is considerably above the Kosterlitz-Thouless transition temperature.  \\

\noindent{{\bf Away from half-filling, and the effect of decreasing $\hbar\omega/E_F$}}\\
Doping away from half filling further suppress the CDW ordering. In the supplementary materials we report the result for $15\%$ electron-doped triangular and square lattices.  Compared with the results at half-filling, SC/CDW are obviously enhanced/suppressed.

Decreasing the phonon frequency makes the Holstein oscillator more classical. Due to the diminished quantum fluctuations,  bipolarons are easier to form and localize. As a result CDW correlation  gets stronger and SC becomes weaker.\\

\noindent{{\bf Conclusion}}\\
We have studied the effects of frustrating charge order at large electron-phonon coupling for the Holstein model through  sign-problem-free QMC simulation. We conclude that frustrating the charge order enables superconductivity to exist under much wider conditions in temperature, electron-phonon coupling strength, and the ratio between the phonon energy and the electron bandwidth. In particular, frustrating the charge order enables a novel coexistence phase where superconducting coherence develops in the presence of charge order. In conclusion, frustrating the charge order formation is a powerful way to enhance superconductivity!

\noindent{{\bf Acknowledgement}} \\

This work was primarily funded by the U.S. Department of Energy, Office of Science, Office of Basic Energy Sciences, Materials Sciences and Engineering Division under Contract No. DE-AC02-05-CH11231 (Theory of Materials  program KC2301). Z.X.L. and D.H.L. also acknowledge support from the Gordon and Betty Moore Foundation's EPIC initiative, Grant GBMF4545. M.L.C. also acknowledges support from the National Science Foundation Grant No. DMR-1508412. Computational resources are provided by the National Energy Research Scientific Computing Center (NERSC).

\begin{widetext}
\section{Supplementary Information}

\renewcommand{\theequation}{S\arabic{equation}}
\setcounter{equation}{0}
\renewcommand{\thefigure}{S\arabic{figure}}
\setcounter{figure}{0}
\renewcommand{\thetable}{S\arabic{table}}
\setcounter{table}{0}

\subsection{I. Details of the Quantum Monte Carlo simulation}

We apply both the finite-temperature and projector determinant QMC algorithm to study \Eq{model}. In the finite-temperature simulation, the grand canonical ensemble averages of observables  are evaluated through: $\avg{\hat{O}} = \Tr[\hat{O}e^{-\beta \hat{H}}]/\Tr[e^{-\beta \hat{H}}]$. Here $\beta$ is the inverse temperature. The  values studied in this paper are $4/t\le\beta\le 16/t$, where $t$ is electron's nearest-neighor hopping matrix element. The imaginary time is discretized with the time step $\Delta\tau=0.1/t$. We have checked that the results do not change upon further decrease of $\Delta\tau$.

In the projector QMC, we evaluate the ground-state expectation values of observables according to $\avg{\hat{O}} =\bra{\psi_0} O \ket{\psi_0}/\<\psi_0 \mid \psi_0\>   = \lim_{\theta\rightarrow \infty} \{\<\psi_T\mid e^{-\theta H } O e^{-\theta H} \mid\psi_T\>/\<\psi_T\mid e^{-2\theta H}\mid \psi_T\>$, where $|\psi_T\>$ is a trial state. In this work $\theta$ is set to $30/t$, and we have checked the convergence of the results against further increase of $\theta$. Like the finite-temperature calculations we have checked that the imaginary time step $\Delta\tau = 0.1/t$ is sufficient to grantee the convergence of the result.

Finally in both the zero and finite temperature calculations we carry out  both single-site and global updates to ensure the statistical independence in our Monte-Carlo sampling.

\begin{figure}[t]
\includegraphics[height=1.8in]{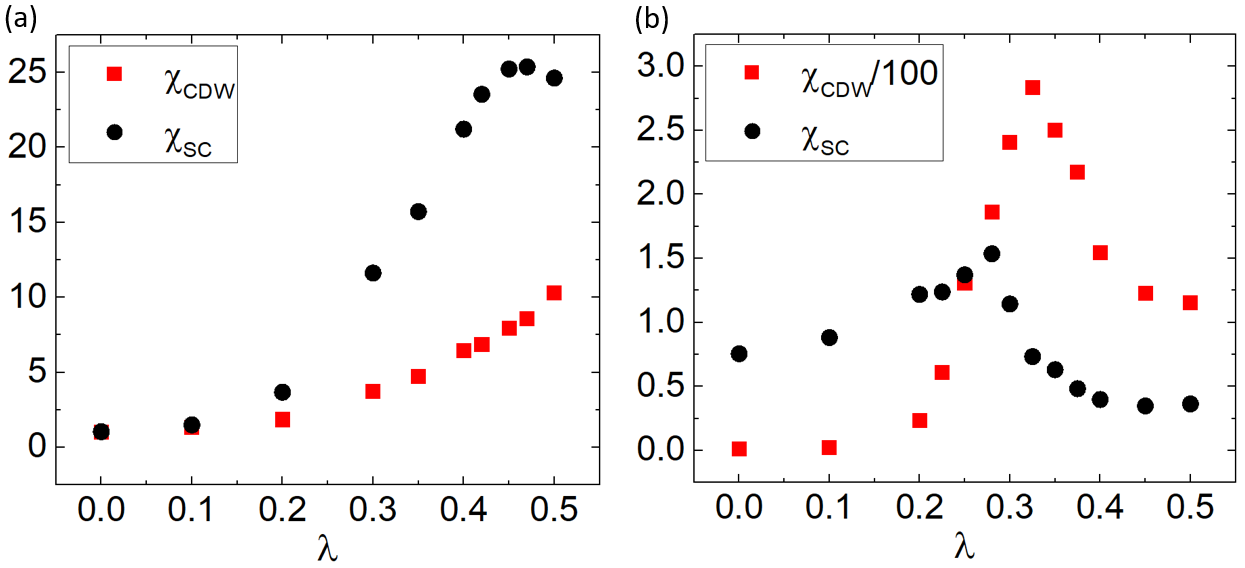}
\caption{(a) The SC and CDW susceptibilities for a doped ($\avg{n}=1.15$) triangular lattice. (b) The SC and CDW susceptibilities for a doped ($\avg{n}=1.15$) square lattice. The results are obtained under the parameter choice $\hbar\omega/E_F= 0.3$, $L=12$ and $T= t/16$. }
\label{figSM1}
\end{figure}

\subsection{II. The non-zero temperature SC and CDW susceptibilities for doped triangular and square lattices}
Intuitively, incommensurate filling factors resulting from doping should suppress the CDW order and  enhance the SC pairing. To check this intuition we calculate the SC and CDW susceptibilities for different values of  $\lambda$ at temperature $T=t/16$ for lattices with linear dimension  $L=12$. The doping is chosen to be 15\%, i.e., $\avg{\hat{n}}=1.15$. The results are shown in \Fig{figSM1}. Compared with half-filling, the CDW susceptibility is suppressed by doping, while the SC susceptibility is enhanced. Moreover by comparing \Fig{figSM1}(a) and \Fig{figSM1}(b) we conclude that for a doped system, lattice frustration remains very effective in suppressing/enhancing CDW/SC orders.

\subsection{III. The non-zero temperature SC and CDW susceptibilities for a triangular lattice at   $\hbar\omega/E_F=0.1$.}
Here the temperature is set to $T=t/16$ and linear system size is $L=12$. In \Fig{figSM2} we show the SC and CDW susceptibilities as a function of $\lambda$. Qualitatively the behaviors of the SC and CDW  susceptibilities are similar to those for $\hbar\omega/E_F=0.3$. However, it is notable that lower $\hbar\omega/E_F$ enhances the CDW  while suppress the SC ordering tendencies.

\begin{figure}[t]
\includegraphics[height=1.8in]{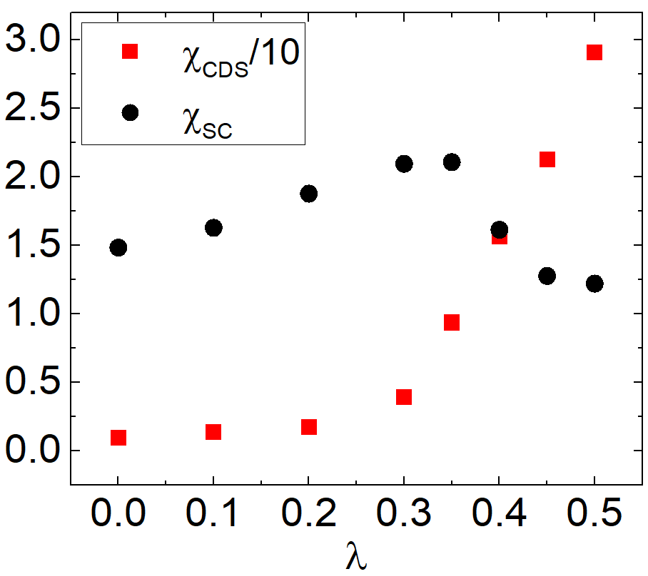}
\caption{The SC and CDW susceptibility on a triangular lattice for $\hbar\omega/E_F =0.1$. The  temperature is $T=t/16$ and the linear system size $L=12$.  }
\label{figSM2}
\end{figure}

\subsection{IV. Finite-size scaling analysis of the zero temperature SC and CDW orders for  triangular and square lattices}
We perform projector QMC simulation to study the ground-state properties of the Holstein model on the triangular lattice.
The Fourier transforms of the SC and CDW correlation functions at momentum $(0,0)$ (SC) and $(4\pi/3,0)$(CDW) are shown in \Fig{figSM3}. When extrapolated to $L=\infty$ the finite values of these quantities indicates long-range order. In panels (a) to (d) the data are fit by second-order polynomials in $1/L$.  At  $\hbar\omega/E_F = 0.1$, the SC order (panel(a)) is persistent in the entire range of $\lambda$ ($0.2\leq \lambda \leq 0.8$). However, the CDW structure factors (panel(b)) are extrapolated to zero or negative within error bar when $\lambda <0.4$, while extrapolated to finite values when $\lambda >0.4$. The result suggests that the ground state is a coexistent phase of SC and $\sqrt{3}\times\sqrt{3}$ CDW when e-ph coupling is stronger than $0.4$. For $\hbar\omega/E_F = 0.3$ the SC correlations (panel(c)) also extrapolate to a finite values in the entire range of $\lambda$ we studied. The CDW correlations (panel(d)), on the other hand, are extrapolated to finite values only when $\lambda >0.6$. To verify this result, we also calculated the RG-invariant ratio $R =S(Q)/S(Q+\delta q) $ as a function of lattice size, where $Q$ is peak momentum of Fourier transformed correlation function, and $\delta q$ is the minimum  allowed momentum quantum on lattice. $R(L)$ has smaller finite-size scaling corrections than correlation functions, hence is a powerful tool for investigating the thermal or quantum phase and phase transition on finite lattices. In the long-range ordered phase, $R(L)$ should diverge for $L \rightarrow \infty$, while $R(L) \rightarrow 0$ for $L \rightarrow \infty$ in disordered phase. At the critical point, $R(L)$ collapses to a finite value for different $L$ due to scaling invariance . We present the results of RG-invariant ratio for $\hbar\omega/E_F=0.1$ in \Fig{fig2}(c),(d) and $\hbar\omega/E_F=0.3$ in \Fig{figSM3}. The results are qualitatively consistent with the conclusion drawn from the extrapolation of the correlation functions. For $\hbar\omega/E_F=0.1$, we observe a quantum phase transition from the SC phase to a SC and CDW coexistent phase around $\lambda \approx 0.42$. For $\hbar\omega/E_F= 0.3$, the CDW order coexists with SC when $ 0.6\lesssim \lambda \leq 0.8$.

\begin{figure}[t]
\includegraphics[height=3.0in]{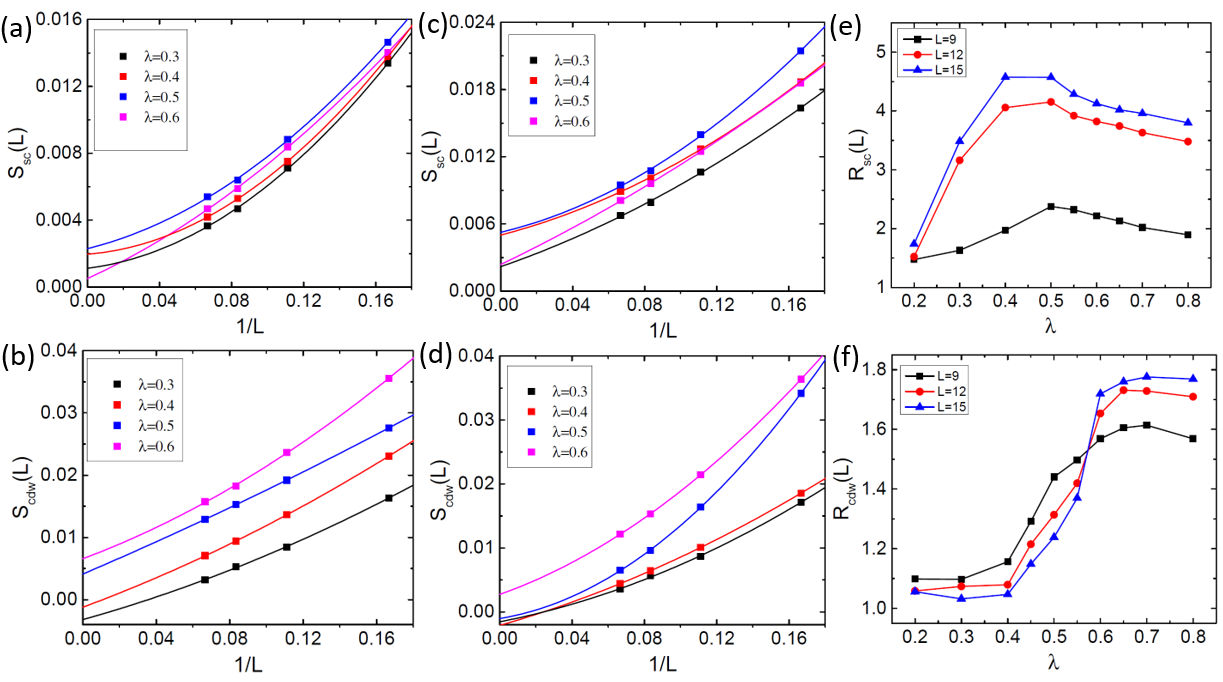}
\caption{(a) Finite size scaling analysis of the peak of the Fourier-transformed SC correlation function on a triangular lattice for $L=6,9,12,15$ and $\hbar\omega/E_F=0.1$. (b) Finite size scaling analysis for the peak of the CDW structure factors on a triangular lattice for $L=6,9,12,15$ and $\hbar\omega/E_F=0.1$. (c) The same as (a) for $\hbar\omega/E_F=0.3$. (d) The same as (b) for $\hbar\omega/E_F=0.3$. (e)  The RG invariant ratio $R_{SC}(L)$ on triangular lattice for $\hbar\omega/E_F=0.3$. (f) The $R_{SC}(L)$ on triangular lattice for $\hbar\omega/E_F=0.3$. In this figure, the temperature is $t/16$ and linear system size $L=12$.  }
\label{figSM3}
\end{figure}

For comparison, we have also studied the Holstein model on square lattice. In this study  we turn on a next nearest neighbor hopping $t_2 = -0.2 t_1$ to get rid of  Fermi surface nesting. We perform finite-size scaling analysis of the SC and CDW correlation functions for $\hbar\omega/E_F=0.1$ and $0.3$. For $ 0.2 \leq \lambda \leq 0.8$. The results of RG invariant ratio, as plotted in \Fig{figSM4}, clearly show that the ground state possesses CDW long-range order and no SC order for both $\hbar\omega/E_F=0.1$ and $0.3$. This result, combined with those for the triangular lattice, suggests that CDW order is strongly suppressed by geometric frustration, which enables SC pairing to exist in a larger range of e-ph coupling strength.

\begin{figure}[t]
\includegraphics[height=3.0in]{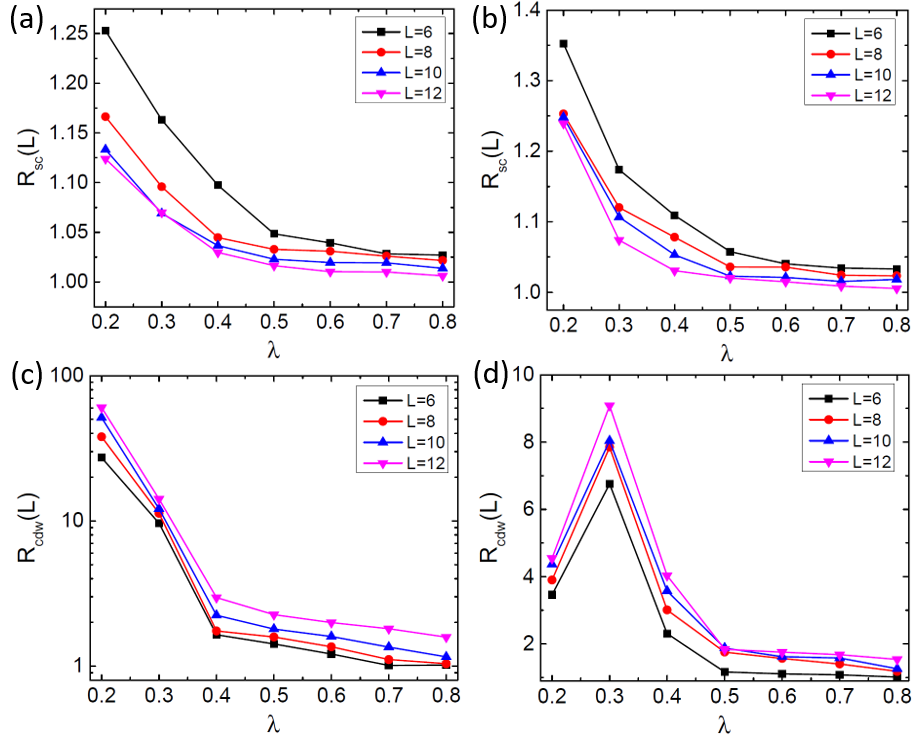}
\caption{The RG invariant ratios $R_{SC}(L)$ and $R_{CDW}(L)$ for the square lattice. (a) The $R_{SC}(L)$  at $\hbar\omega/E_F=0.1$, and (b) the $R_{SC}(L)$ at $\hbar\omega/E_F=0.3$. (c) The $R_{CDW}(L)$ at $\hbar\omega/E_F=0.1$, and (d) the $R_{CDW}(L)$ at $\hbar\omega/E_F=0.3$. }
\label{figSM4}
\end{figure}

\subsection{V. Local density and density fluctuation distribution in SC and CDW coexistent phase}
We compute the average of the charge density and its mean square fluctuation in the SC and CDW coexistent phase. We employ projector QMC for $\lambda=0.5$ and $\hbar\omega/E_F 0.1$. A tiny modulated chemical potential consistent with CDW periodicity is added to pin the CDW to one of three degenerate CDW patterns. In particular, we applied a modulated chemical potential with amplitude $\delta=0.02$ and have checked that such pinning potential do not affect the intrinsic values of the SC and CDW correlation functions. The result of averaged charge density is shown in \Fig{fig4}(a), which clearly reveals the $\sqrt{3}\times\sqrt{3}$ CDW pattern in \Fig{fig1}(a). The charge density on the three sublattices are $(1+a,1,1-a)$. More importantly, as shown in \Fig{fig4}(b), the charge fluctuations are significantly stronger on the sublattice where the averaged density is approximately unity. Since SC order requires charge fluctuation, we expect that SC correlation to be bigger on such sulattice. This is verified by our unbiased QMC simulation.

\subsection{VI. Electron spectral function on triangular and square lattice}
 In order to investigate the existence of pseudogap in the Holstein model, we compute the electron spectral function through stochastic analytical continuation of the imaginary-time Green's function. We obtain the electron spectral functions at different points on the Fermi surface. Here we present the spectral function $A(k_F,\omega)$ at momentum point where the pseudogap is the minimum. The results of $A(k_F,\omega)$ for $\lambda=0.4$ on triangular at temperatures $T=t/18,t/12,t/8$ are shown in \Fig{figSM5} where $t/8$ is above the $T_c$ for triangular lattice (the square lattice does not show a SC transition).  We estimate the value of single-particle gap from the peaks of spectral function $A(k_F,\omega)$. In \Fig{fig5}, we present the values of spectral gap for several values of $\lambda$ on triangular and square lattices. The pseudogap above $T_c$ undergos a sharp upturn around $\lambda$ value where CDW susceptibility rises. We also present the single-particle gap as a function of temperature for $\lambda=0.4$ on triangular in \Fig{figSM5}(d). From this result we estimate the onset temperature of pseudogap $T^* \approx 0.45 t$.

\begin{figure}[t]
\includegraphics[height=3.0in]{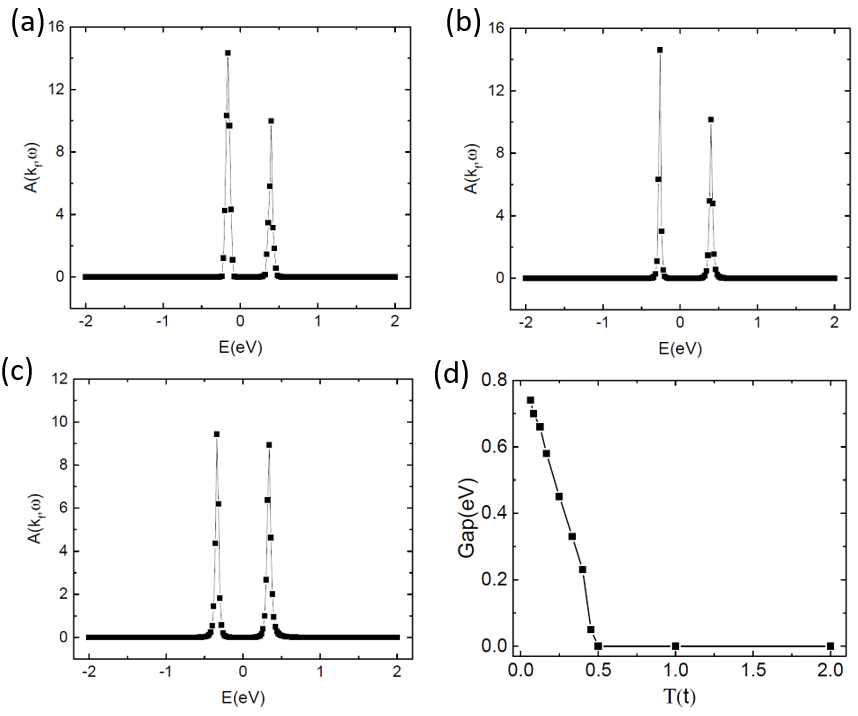}
\caption{ The electron spectral function $A(k_F,w)$ for $\lambda=0.4$ on triangular lattice at various temperatures: (a) T= t/8; (b) T=t/12; (c) T=t/16. The results clearly show that spectral gap survives above the SC transition temperature $t/10$. (d) The single-particle gap on triangular lattice for  $\lambda=0.4$ as a function of temperature. The pseudogap onset temperature is estimated to be $T^*\approx 0.45 t.$  }
\label{figSM5}
\end{figure}

\end{widetext}

\end{document}